\documentclass[12pt]{article}
\usepackage{graphicx}% Include figure files
\usepackage{subfigure}
\usepackage{naturenew}
\usepackage{naturefem}
\usepackage{citesupernumber}
\usepackage{nature}

\begin{document}

\bibliographystyle{nature}

\title{Geometry and symmetry presculpt the free-energy landscape of proteins}

\author{Trinh Xuan Hoang$^1$, Antonio Trovato$^2$, Flavio Seno$^2$, \\
Jayanth R. Banavar$^3$ and Amos Maritan$^{2,4}$}

\date{}

\maketitle

\noindent
$^1$Institute of Physics, Natural Center for Natural Science and Technology, 
46 Nguyen Van Ngoc, Hanoi, Vietnam

\noindent
$^2$INFM and Dipartimento di Fisica `G. Galilei',
Universit\`a di Padova, Via Marzolo 8, 35131 Padova, Italy

\noindent
$^3$Department of Physics, 104 Davey Lab, The Pennsylvania State University,
University Park PA 16802, USA

\noindent
$^4$The Abdus Salam International Center for Theoretical Physics
(ICTP), Strada Costiera 11, 34014 Trieste, Italy

%\noindent $^5$ International School for Advanced Studies (SISSA) and INFM, Via
%Beirut 2-4, 34014 Trieste, Italy

\vspace{4cm}

\noindent {\bf Correspondence} and requests for materials should
be addressed to JRB (banavar@psu.edu) or to AM (maritan@pd.infn.it).

\newpage
\begin{abstract}
We present a simple physical model which demonstrates that the native
state folds of proteins can emerge on the basis of considerations of
geometry and symmetry. We show that the inherent anisotropy of a chain
molecule, the geometrical and energetic constraints placed by the
hydrogen bonds and sterics, and hydrophobicity are sufficient to yield
a free energy landscape with broad minima even for a homopolymer.
These minima correspond to marginally compact structures comprising
the menu of folds that proteins choose from to house their
native-states in. Our results provide a general framework for
understanding the common characteristics of globular proteins.

\end{abstract}

%\newpage
\newcounter{ctr}
\setcounter{ctr}{1}

Protein folding\cite{Funnel2,Funnel3,Funnel4,Fersht,B3} is complex
because of the sheer size of protein molecules, the twenty types of
constituent amino acids with distinct side chains and the essential
role played by the environment.  Nevertheless, proteins fold into a limited
number\cite{ChFin90,Chothia1} of evolutionarily conserved
structures\cite{Denton,Chothia2}.
It is a familiar, yet remarkable,
consequence of symmetry and geometry that ordinary
matter crystallizes in a limited number of distinct forms.  Indeed,
crystalline structures transcend the specifics of the various
entities housed in them.   Here we ask the analogous
question\cite{Wol96}: is the menu  of protein folds
also determined by geometry and symmetry?

We show that a simple model which encapsulates a few general
attributes common to all polypeptide chains, such as steric
constraints\cite{Rama,LINUS,Baldwin}, hydrogen
bonding\cite{Pauling1,Pauling2,Eisen} and
hydrophobicity\cite{Kauzmann}, gives rise to the emergent free energy
landscape of globular proteins.  The relatively few minima in the
resulting landscape correspond to putative marginally-compact
native-state structures of proteins, which are assemblies of helices,
hairpins and planar sheets.  A superior fit\cite{Frustration,Brenner}
of a given protein or sequence of amino acids to one of these
pre-determined folds dictates the choice of the topology of its
native-state structure. Instead of each sequence shaping its own free
energy landscape, we find that the overarching principles of geometry
and symmetry determine the menu of possible folds that the sequence
can choose from.

Following Bernal\cite{Bernal},
the protein problem  can be divided into two distinct
steps: first, analogous to the elucidation of crystal structures, one must
identify the essential features that account for
the common characteristics of all proteins; second, one must understand
what makes one protein different from another. Guided by recent
work\cite{Tubone,RMP} which has shown that a faithful description of a
chain molecule is a tube and using information from known protein
native state structures, our focus, in this paper, is on the first step
-- we demonstrate that the native-state folds of proteins emerge from
considerations of symmetry and geometry within the context of a simple model.

We model a protein as a chain of {\em identical} amino acids,
represented by their $C_{\alpha}$ atoms, lying along the axis of a
self-avoiding flexible tube.  The preferential parallel placement of nearby
tube segments approximately mimics the effects of
the anisotropic interaction of hydrogen bonds,
while the space needed for the clash-free packing of side chains is
approximately captured by the non-zero tube thickness\cite{Tubone,RMP}.
Here we carefully incorporate these key geometrical features via an
extensive statistical analysis of experimentally determined native
state structures in the Protein Data Bank (PDB).

A tube description places constraints on the radii of circles drawn through
both local and non-local triplets of  $C_{\alpha}$ positions of a
protein native structure\cite{RMP,BMMT02}. Furthermore, when one deals with a
chain molecule, the tube picture underscores the crucial importance of knowing
the context that an amino acid is in within the chain. The standard
coarse-grained approach considers the locations of interacting amino acid
pairs. Here, instead, we incorporate the strongly directional hydrogen bonding
between a pair of amino acids, through an analysis of the PDB to
determine the constraints on the mutual orientation of the local coordinate
systems defined from a knowledge of the locations of the
$C_{\alpha}$ atoms (see Methods and Figure 1).
The geometrical constraints associated with the
tube and hydrogen bonds, that we consider here,
are representative of the typical {\em aspecific}
behavior of the interacting amino acids.

There are two other ingredients in the model: a local bending penalty which is
related to the steric hindrance of the amino acid side chains and a pair-wise
interaction of the standard type mediated by the water\cite{Kauzmann}. Even
though these two properties clearly depend on the {\em specific} amino acids
involved in the interaction, here we choose to study the phase diagram of a
{\em homo}-peptide chain by varying its overall hydrophobicity and local
bending penalty, while keeping them constant along the chain. This is the
simplest and most general way to assess their relevance in shaping the
free-energy landscape.

\section*{Methods}

\noindent
{\bf Tube geometry}.
The protein backbone is modeled as a chain of
$C_{\alpha}$ atoms (Figure 2a) with a fixed distance of $3.8 \AA$ between
successive atoms along the chain, an excellent assumption for all but
non-cis Proline amino acids\cite{Creighton}. The geometry imposed by
chemistry dictates that the bond angle associated with three
consecutive $C_{\alpha}$ atoms is between $82^{\circ}$ and
$148^{\circ}$.

Self-avoiding conformations of the tube whose axis is the protein
backbone are identified by considering all triplets of $C_{\alpha}$
atoms and drawing circles through them and ensuring that none of their
radii is smaller than the tube radius\cite{JSP} (Figure 2a).  At the
local level, the three body constraint ensures that a flexible tube
cannot have a radius of curvature any smaller than the tube thickness
in order to prevent sharp corners whereas, at the non-local level, it
does not permit any self-intersections.  There is an inherent local
anisotropy due to the special direction singled out by consecutive
atoms along the chain which enforces a preference for parallel
alignment of neighboring tube segments in a compact conformation.

The backbone of $C_{\alpha}$ atoms is treated as a flexible tube of
radius $2.5 \AA$, a constraint imposed on all (local and non-local)
three body-radii, an assumption validated for protein native
structures\cite{BMMT02}.  It is interesting to note that recent
observations of residual dipolar couplings in short
peptides\cite{Shortle} in the denatured state have demonstrated their
stiffness and their anisotropic deformability -- the building blocks
of proteins are relatively stiff segments with strong directional
preferences.
\vspace{20pt}

\noindent
{\bf Sterics}. Steric constraints require that no two non-adjacent
$C_{\alpha}$ atoms are allowed to be at a distance closer than $4
\AA$. Ramachandran and Sasisekharan\cite{Rama} showed that steric
considerations based on a hard sphere model lead to clustering of the
backbone dihedral angles in two distinct $\alpha$ and $\beta$ regions
for non-glycyl and non-prolyl residues. The two backbone geometries
that allow for systematic and extensive hydrogen
bonding\cite{Pauling1,Pauling2,Eisen} are the $\alpha$-helix and the
$\beta$-sheet obtained by a repetition of the backbone dihedral angles
from the two regions respectively\cite{Baldwin}. Short chains rich in
alanine residues, which are a good approximation to a stretch of the
backbone, can adopt a helical conformation in water (see
\cite{Ing,Marq,Scholtz,Spek,Vila1,Vila2} for a detailed discussion
of experimental conditions that would lead to a helical conformation). 
However, when
one has more heterogeneous side chains, the helix backbone could
sterically clash with some side chain conformers resulting in a loss
of conformational entropy\cite{Cream}. When the price in side chain
entropy is too large, an extended backbone conformation results
pushing the segment towards a $\beta$-strand
structure\cite{Baldwin}. These steric constraints are approximately
imposed through an energy penalty (denoted by $e_R$) when the local
radius of curvature is between $2.5 \AA$ and $3.2 \AA$. (The magnitude
of the penalty does not depend on the specific value of the radius of
curvature provided it is between these values.) There is no cost when
the local radius exceeds $3.2 \AA$.  Note that the tube constraint
does not permit any local radius of curvature to take on a value less
than the tube radius, $2.5 \AA$.
\vspace{20pt}

\noindent
{\bf Hydrogen bonds}. We do not allow more than two
hydrogen bonds to form at a given $C_{\alpha}$ location.
In our representation of the protein backbone,
local hydrogen bonds form between $C_{\alpha}$ atoms separated by
two residues along the sequence  with an energy defined to be $-1$ unit,
whereas non-local hydrogen bonds are those that form between
$C_{\alpha}$ atoms separated by more than three residues along the sequence
with an energy of $-0.7$.
This energy difference is based on
experimental findings that the local bonds provide more
stability to a protein than do the non-local hydrogen
bonds\cite{Sosnick}. Cooperativity effects\cite{Liwo,Fain} are taken into
account by adding an energy of $-0.3$ units when consecutive hydrogen bonds
along the sequence are formed.
There is some latitude in the choice of the values of these energy parameters.
The results that we present are robust to changes (at least of the
order of 20\%) in these parameters.

\vspace{20pt}
\noindent
{\bf Geometrical constraints due to hydrogen bonding}.
Three non-collinear consecutive atoms
($i-1$,$i$,$i+1$) of the chain define a plane. At atom $i$ (special
care is needed to adapt these rules to atoms at the $C$ and
$N$-termini), one may define a tangent vector (along the direction
joining the $i-1$ and $i+1$ atoms) and a normal vector (along the
direction joining the $i$-th atom and the center of the circle passing
through the three atoms), which together define a plane.  One then
defines a binormal vector $\vec b_i$ perpendicular to the plane with
the tangent, normal and binormal forming a right-handed local
coordinate system (Figure 1).  This coordinate system defines the context of an
amino acid within a chain, a feature that plays a crucial role in the
tube picture.  For hydrogen bond formation between atom $i$ and $j$,
the distance between
these atoms ought to be between $4.7 \AA$ and $5.6 \AA$ ($4.1 \AA$ and
$5.3 \AA$) for the local (non-local) case. A study of
protein native state structures reveals an overall nearly parallel
alignment of the axes defined by three vectors: the binormal vectors
at $i$ and $j$ and the vector $\vec r_{ij}$ joining the $i$ and $j$
atoms.  A hydrogen bond is allowed to form only when the binormal axes
are constrained to be within $37^{\circ}$ of each other, whereas the
angle between the binormal axes and that defined by $\vec r_{ij}$
ought to be less than $20^{\circ}$.  Additionally, for
the cooperative formation of non-local hydrogen bonds, one requires
that the corresponding binormal vectors of successive $C_{\alpha}$
atoms make an angle greater than $90^{\circ}$.  The
first and the last residues of the chain are special cases since their
binormal vectors are not defined. In order for such residues to form a
hydrogen bond (with each other or with other internal residues in the
chain), it is required that the angle between the associated ending
peptide link and the connecting vector to the other residue
participating in the hydrogen bond is between $70^{\circ}$ and
$110^{\circ}$. As in real protein structures, when helices are formed,
they are constrained to be right-handed.  This constraint is enforced by
requiring that the backbone chirality associated with each local
hydrogen bond is positive. The chirality is defined as the sign of the
scalar product $({\vec r}_{i,i+1} \times {\vec r}_{i+1,i+2}) \cdot
{\vec r}_{i+2,i+3}$.

Our approach for the derivation of the geometrical constraints imposed
by hydrogen bonds is similar to that carried out at the level of an
all-atom description of the protein chain\cite{Cohen}.  For the
simpler $C_{\alpha}$ atom based description, hydrogen bond energy
functions have been introduced previously\cite{Jeff,Jesper} but
without any input from a statistical analysis of protein structures.

\vspace{20pt}
\noindent
{\bf Hydrophobic interactions}. The hydrophobic (hydrophilic) effects
mediated by the water are captured through a relatively weak
interaction, $e_W$, (either attractive or repulsive) between
$C_{\alpha}$ atoms which are within $7.5 \AA$ of each other
(Figure 2c).  Note that
hydrogen bonds can easily be formed between the amino acid residues in
an extended conformation and the water molecules.
Within our model, the
intrachain hydrogen bond interaction introduces an effective
attraction, because water molecules are not explicitly present. The
hydrophobicity scale is thus renormalized (e.g. even when
$e_W$ is weakly positive, there could be an effective
attraction resulting in structured conformations such as a single helix or
a planar sheet). A
negative $e_W$ is, in any case, crucial for promoting the assembly of
secondary motifs in native tertiary arrangements.
The properties of the model are summarized in Table 1.

\section*{Results and Discussion}

Figure 3 shows the ground state phase diagram obtained from Monte-Carlo computer
simulations using the simulated annealing technique\cite{Anneal}.
(The solvent-mediated energy, $e_W$, and the local radius of
curvature energy penalty, $e_R$, (see Methods for a description of the energy
parameters) are measured in units of the local hydrogen bond energy.)
When $e_W$
is sufficiently repulsive (hydrophilic) (and $e_R > 0.3$ in the phase diagram),
one obtains a swollen phase with very few contacts between the $C_{\alpha}$
atoms. When $e_W$ is sufficiently attractive, one finds a very compact,
globular phase with featureless ground states with a high number of
contacts.

Between these two phases (and in the vicinity of the swollen phase), a
marginally compact phase emerges (the interactions barely stabilize
the ordered phase) with distinct structures including a single helix,
a bundle of two helices, a helix formed by $\beta$-strands, a
$\beta$-hairpin, three-stranded $\beta$-sheets with two distinct
topologies and a $\beta$-barrel like conformation.  Strikingly, these
structures are the stable ground states in different parts of the
phase diagram.  Furthermore, other conformations, closely resembling
distinct super-secondary arrangements observed in
proteins\cite{ChFin90}, such as the $\beta$-$\alpha$-$\beta$ motif,
are found to be competitive local minima, whose stability can be
enhanced by sequence design (for example, non-uniform values of
curvature energy penalties for single amino acids and hydrophobic
interactions for amino acid pairs).  Figure 4 shows a compendium of
various structures obtained in our studies including, for comparison,
a generic compact conformation of a conventional polymer chain (with
no tube geometry or hydrogen bonds), which neither is made up of
helices or sheets nor possesses the significant advantages of protein
structures. While there is a remarkable similarity between the
structures that we obtain and protein folds, our simplified
coarse-grained model is not as accurate as an all-atom representation
of the poly-peptide chain in capturing features such as the packing of
amino acid side chains.

The fact that different putative native structures are found to be
competing minima for the same homopolymeric chain clearly establishes
that the free-energy landscape of proteins is pre-sculpted by means of
the few ingredients utilized in our model. At the same time,
relatively small changes in the parameters $e_W$ and $e_R$ lead to
significant differences in the emergent ground state structure,
underscoring the sensitive role played by chemical heterogeneity in
selecting from the menu of native state folds.

Figure 5a is a contour plot of the free energy at a temperature higher
than the folding transition temperature (identified by the specific
heat peak) for the parameter values $e_W=-0.08$ and $e_R=0.3$ for
which the ground state is an $\alpha$-helix (Figure 3). The free
energy landscape has just one minimum corresponding to the denatured
phase whose typical conformations are somewhat compact but
featureless.  The contour plot at the folding transition temperature
(Figure 5b) has three local minima corresponding to an $\alpha$-helix,
a three-stranded $\beta$-sheet and the denatured state. At lower
temperatures, the $\alpha$-helix is increasingly favored and the
$\beta$-sheet is never the global free energy minimum.  Many protein
folding experiments show that for small globular proteins, at the
transition temperature, only two states (folded and unfolded) are
populated. The appearance in the present framework of multiple states
for a homopolymer chain suggests that two state folders might have
been evolutionarily selected by sequence design favoring the native
state conformation over competing folds in the pre-sculpted landscape.

Such a design is indeed straightforward within our model.  For
example, the $\alpha$-$\beta$-$\alpha$ motif shown in Figure 4d (which
is a local energy minimum for a homopolymer) can be stabilized into a
global energy minimum for the sequence HPHHHPPPPHHPPHHPPPPHHHPP, with
$e_W=-0.4$ for HH contacts and $e_W=0$ for other contacts, and
$e_R=0.3$ for all residues.

It is interesting to note that lattice models of compact homopolymers
yield large amounts of secondary structure\cite{Chan}--local radius of
curvature constraints are built into lattice models.
However, an all atom study of poly-alanine has shown that
compactness alone is insufficient to obtain secondary
structures\cite{Chan2}. Even a simple tube subject to an attractive
self-interaction favoring compaction has a tendency to form helices,
hairpins and sheets when the ratio of the tube thickness to the range
of attractive interaction is tuned properly\cite{RMP}. Our work here
underscores the importance of hydrogen bonds in stabilizing both
helices and sheets simultaneously (without any need for adjustment of
the tube thickness) allowing the formation of tertiary arrangements of
secondary motifs. Indeed, the fine-tuning of the hydrogen bond and the
hydrophobic interaction is of paramount importance in the selection of
the marginally compact region of the phase diagram in which protein
native folds are found. It is also important to note that proteins are
relatively short chain molecules compared to conventional
polymers. These are special features of proteins, which distinguishes
them from generic compact polymers.

A free energy landscape with a 1000 or so minima\cite{Chothia1} with
correspondingly large basins of attraction leads to stability and
diversity, the dual characteristics needed for evolution to be
successful.  Proteins are those sequences which fit
well\cite{Frustration} into one of these minima and are relatively
stable.  Yet, the fact that the marginally compact phase lies in the
vicinity of a phase transition to the swollen phase allows for an
exquisite sensitivity of protein structures to the right types of
perturbations.  Thus a change in the external environment (e.g. an ATP
molecule binding to the protein) could reshape the free energy
landscape allowing for a different, stable and easily foldable
conformation.

In summary, within a simple, yet realistic, framework, we have shown
that protein native-state structures can arise from considerations of
symmetry and geometry associated with the polypeptide chain. The
sculpting of the free energy landscape with relatively few broad
minima is consistent with the fact that proteins can be designed to
enable rapid folding to their native states. The limited number of folds
arises from the geometrical constraints imposed by sterics and
hydrogen bonds. In the marginally compact phase, not only does one
have a space-filling conformation (the nearby backbone segments have
to be placed near each other in order to avail of the attractive
potential), which is effective in expelling water from the hydrophobic
core, but also these segments need to have the right orientation with
respect to each other in order to respect the geometrical constraints
imposed by the hydrogen bonds.

\bibliography{pfold_pnas}

\vspace{1cm}

\noindent {\bf Acknowledgements}
We thank Buzz Baldwin, Hue-Sun Chan, Morrel Cohen, Russ Doolittle, Davide
Marenduzzo, George Rose, and Harold Scheraga for their invaluable comments.
This work was supported by Progetti Di Rilevante Interesse Nazionale 2003,
Fondo Integrativo Speciale Ricerca 2001, the National Aeronautics and Space
Administration, National Science Foundation (NSF) Integrative Graduate Education
and Research Traineeship Grant DGE-9987589, NSF Materials Research Science
and Engineering Centers, and the award of a postdoctoral
fellowship at the Abdus Salam International Center for Theoretical Physics
(to T.X.H.).

%\vspace{1cm}

%\noindent {\bf Competing interests statement} The authors declare
%that they have no competing financial interests.

%\vspace{1cm}

\newpage
%\section*{Tables}
\subsection*{Table 1. Properties of the model}
\begin{center}
\begin{tabular}{ll}
Parameter & Constraint  \\
\hline
Tube approximation $^{(a)}$ & $R_{ijk} \geq 2.5\AA$, $\forall i<j<k$ \\
local radius of curvature & $2.5\AA \leq R_{i-1,i,i+1}\leq 7.9\AA, \; \forall 1<i<N ^{(b)}$ \\
self avoidance & $r_{ij} \geq 4\AA, \; \forall i<j-1$ \\
amino acid specific? & no \\
%obtained from?  & PDB \\
\hline
Local hydrogen bond $^{(c)}$ & $j=i+3$ \\
$C_\alpha$-$C_\alpha$ distance & $4.7\AA \leq r_{ij} \leq 5.6\AA$ \\
binormal-binormal correlation$^{(d)}$& $|\vec{b}_i \cdot \vec{b}_j| > 0.8$\\
binormal-connecting vector$^{(d,e,f)}$ & $|\vec{b}_i \cdot \vec{c}_{ij}| > 0.94$,
$|\vec{b}_j \cdot \vec{c}_{ij}| > 0.94$ \\
chirality & $(\vec{r}_{i,i+1} \times \vec{r}_{i+1,i+2}) \cdot \vec{r}_{i+2,i+3} > 0 $ \\
energy & $-1$ \\
amino acid specific? & no \\
%obtained from?  & PDB \\
\hline
Non-local hydrogen bond $^{(c)}$ & $j>i+4$ \\
$C_\alpha$-$C_\alpha$ distance & $4.1\AA \leq r_{ij} \leq 5.3\AA$ \\
binormal-binormal correlation$^{(d)}$ & $|\vec{b}_i \cdot \vec{b}_j| > 0.8$\\
binormal-connecting vector$^{(d,e)}$ & $|\vec{b}_i \cdot \vec{c}_{ij}| > 0.94$,
$|\vec{b}_j \cdot \vec{c}_{ij}| > 0.94$ \\
energy & $-0.7$ \\
amino acid specific? & no \\
%obtained from?  & PDB \\
\hline
Cooperative hydrogen bonds & between $(i,j)$ and $(i \pm 1,j \pm 1)$\\
$\beta$-sheet zig-zag pattern $^{(d,g)}$ & $\vec{b}_i \cdot \vec{b}_{i\pm1} < 0$, $\vec{b}_j \cdot \vec{b}_{j\pm1} < 0$ \\
energy per pair & $-0.3$ \\
amino acid specific? & no \\
\hline
Bending rigidity & $R_{i-1,i,i+1} \leq 3.2\AA$ \\
energy & $e_R$ \\
amino acid specific? & yes (for a heteropolymer)\\
\hline
Hydrophobic contact & $j>i+2$ \\
$C_\alpha$-$C_\alpha$ distance & $r_{ij} \leq 7.5\AA$ \\
energy & $e_W$ \\
amino acid specific? & yes (for a heteropolymer) \\
\hline
\end{tabular}
\end{center}
{\small $^{(a)}$ $R_{ijk}$ is the radius of a circle drawn through the
$C_\alpha$ positions of $i$, $j$ and $k$ \\ $^{(b)}$ $N$ is the number
of residues \\ $^{(c)}$ each residue is constrained to form no more
than 2 hydrogen bonds (except the residues located at the chain termini
which form at most 1 hydrogen bond) \\
$^{(d)}$ applied only when the corresponding
binormal vectors exist \\ $^{(e)}$ for $i=1$ and(or) $j=N$ this is
replaced by the constraint that the connecting vector is making an
angle between $70^{\circ}$ and $110^{\circ}$ with the extremal peptide
links. \\ $^{(f)}$ the connecting vector, $\vec{c}_{ij} \equiv
\vec{r}_{ij}/r_{ij}$, is a unit vector joining $i$ and $j$ \\ $^{(g)}$
applied when at least one of the two cooperative hydrogen bonds is
non-local }

\section*{Table Legends}

\subsection*{Table 1}

Summary of all geometrical and energetical parameters
involved in the model definition. All geometrical properties
has been derived via a thorough analysis of PDB native
structures.

\newpage

\section*{Figure Legends}

\subsection*{Figure 1}
Sketch of the local coordinate system. For each $C_{\alpha}$ atom $i$
(except the first and the last one), the axes of a right-handed local
coordinate system are defined as follows. The tangent vector $\hat
t_i$ is parallel to the segment joining $i-1$ with $i+i$. The normal
vector $\hat n_i$ joins $i$ to the center of the circle passing
through $i-1$, $i$, and $i+1$ and it is perpendicular to $\hat
t_i$. $\hat t_i$ and $\hat n_i$ along with the three contiguous
$C_{\alpha}$ atoms lie in a plane shown in the figure. The binormal
vector $\hat b_i$ is perpendicular to this plane. The vectors $\hat
t_i$, $\hat n_i$, $\hat b_i$ are normalized to unit length.

\subsection*{Figure 2}

Sketch of a portion of a protein chain.  The black spheres
represent the $C_{\alpha}$ atoms of the amino acids. The local
radius of curvature $r$ is defined as the radius of the circle passing
through three consecutive atoms and is constrained to lie between
$2.5 \AA$ and $7.9 \AA$ ($r_{max}$). A penalty $e_R$ is
imposed when $2.5 \leq r \leq 3.2$ (see (b)). The hydrophobic
interaction, $e_W$, is operative when two atoms separated by more
than two along the sequence are within $7.5 \AA$ of each other (see
(c)). Note that two non-adjacent atoms cannot be closer than $4 \AA$.
A flexible tube is characterized by the constraint that none of
the three-body radii is less than the tube thickness, chosen here
to be $2.5 \AA$ (see (b) and (d)).

\subsection*{Figure 3}

Phase diagram of ground state conformations.

The ground state conformations were obtained using Monte-Carlo
simulations of chains of 24 $C_{\alpha}$ atoms.  $e_R$ and $e_W$
denote the local radius of curvature energy penalty and the solvent
mediated interaction energy respectively.  Over $600$ distinct local
minima were obtained in different parts of
parameter space starting from a random conformation and
successively distorting the chain with pivot and crankshaft moves
commonly used in stochastic chain dynamics\cite{Sokal}.
A Metropolis Monte-Carlo
procedure is employed with a thermal weight
$\exp\left(-E/T\right)$, where $E$ is the energy of the conformation
and the temperature $T$ is set initially at a high value and then
decreased gradually to zero.  In the orange phase, the ground state is
a 2-stranded $\beta$-hairpin.  Two distinct topologies of a 3-stranded
$\beta$-sheet (dark and light blue phases) are found corresponding to
conformations shown in conformations i and j in Fig. 4 respectively. The helix
bundle shown in conformation b in Fig. 4 is the ground state in the green phase
whereas the ground state conformation in the magenta phase has a
slightly different arrangement of helices.  The white region in the
left of the phase diagram has large attractive values of $e_W$ and the
ground state conformations are compact globular structures with a
crystalline order induced by hard sphere packing
considerations\cite{Karplus} and not by hydrogen bonding (conformation
l in Fig. 4).

\subsection*{Figure 4}

MolScript representation of the most common structures obtained in our
simulations.

Helices and strands are assigned when local or non-local
hydrogen bonds are formed according to the described
rules.  Conformations (a), (b), (h), (i), (j), and (k)
are the stable ground
states in different parts of the parameter space shown in
Figure  4. Conformations (c), (d), (e), (f), and (g) are
competitive local minima.
Conformation (l) is that of a generic compact
polymer chain, obtained by switching off
hydrogen bonds, the tube constraint
and curvature energy penalty  and  is obtained on
maximizing the total number of hydrophobic contacts.

\subsection*{Figure 5}

Contour plots of the effective free energy at high temperature ($T =
0.22$) and at the folding transition temperature $T_f=0.2$.

The effective free energy, defined as
$F(N_l+N_{nl},N_W)=-\ln P(N_l+N_{nl},N_W)$, is obtained as a function of
the total number of hydrogen bonds $N_l+N_{nl}$ and the total number
of hydrophobic contacts $N_W$ from the histogram $P(N_l+N_{nl},N_W)$
collected in equilibrium Monte-Carlo simulations at constant
temperature. The spacing between consecutive levels in each contour
plot is $1$ and corresponds to a free energy
difference of  $k_B\tilde{T}$, where $\tilde{T}$ is the temperature
in physical units. The darker the color, the lower the free energy value.
There is just one free energy
minimum corresponding to the denatured state at a temperature
higher than the folding transition temperature  (Panel (a)) whereas
one can discern the existence of three distinct minima at the
folding transition temperature (Panel (b)).
Typical conformations from each of the minima are shown
in the figure.

\newpage
\pagestyle{empty}

\begin{figure}
\centering
\includegraphics[width=10.4cm,angle=-90]{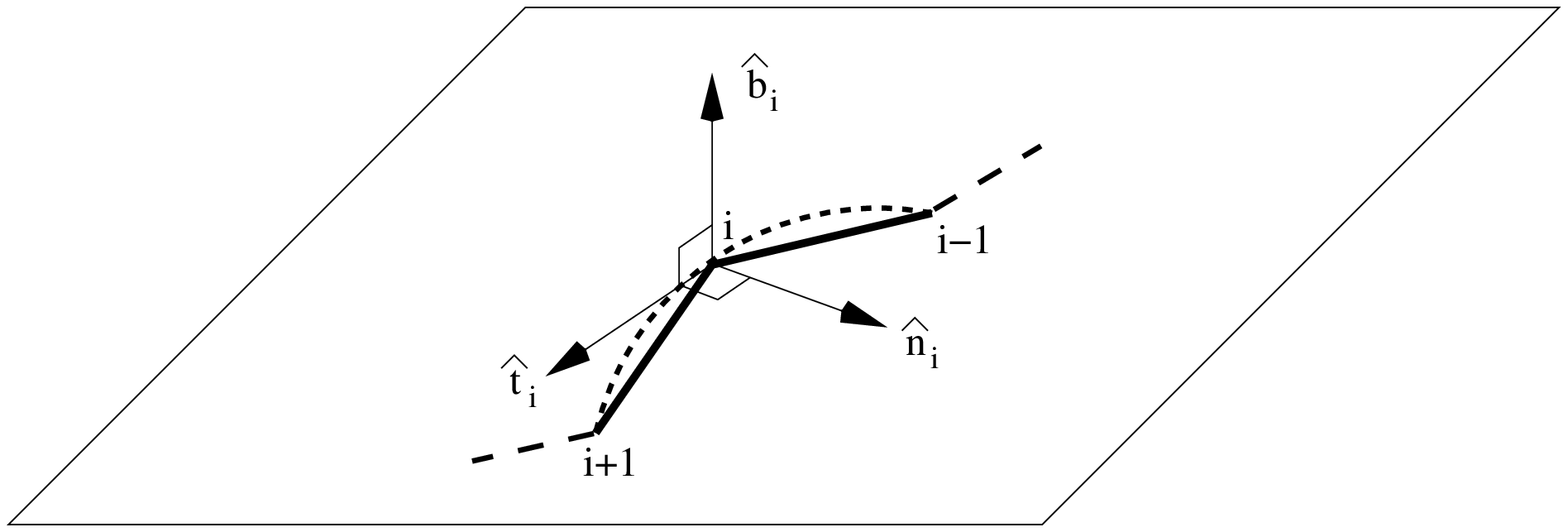}
%\caption{}
\end{figure}
\centerline{Figure-1 (Banavar)}

\newpage
\pagestyle{empty}

\begin{figure}
\centering
\includegraphics[width=10.4cm,angle=-90]{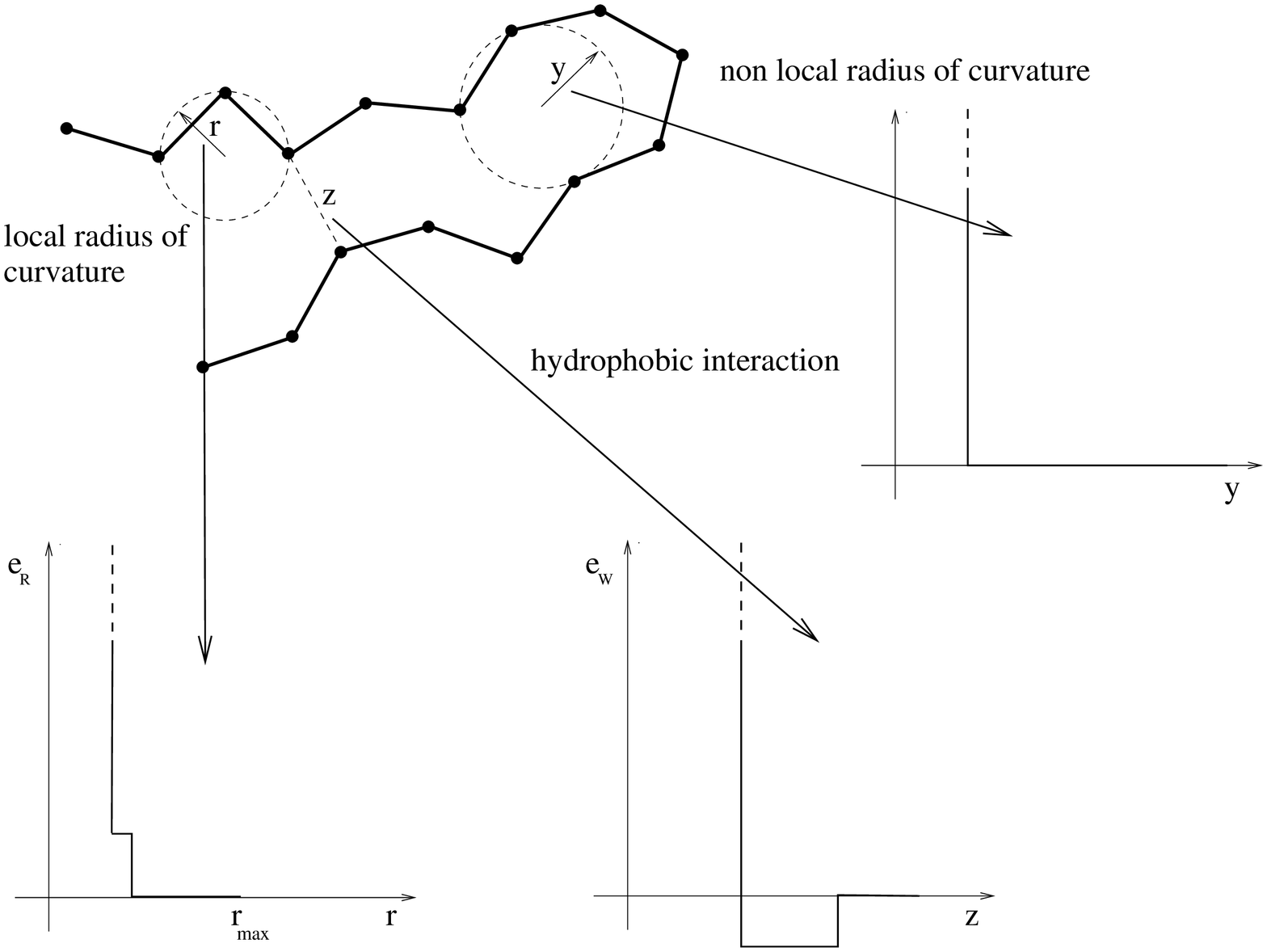}
%\caption{}
\end{figure}
\centerline{Figure-2 (Banavar)}

\newpage
\pagestyle{empty}

\begin{figure}
\centering
\includegraphics[width=10.4cm]{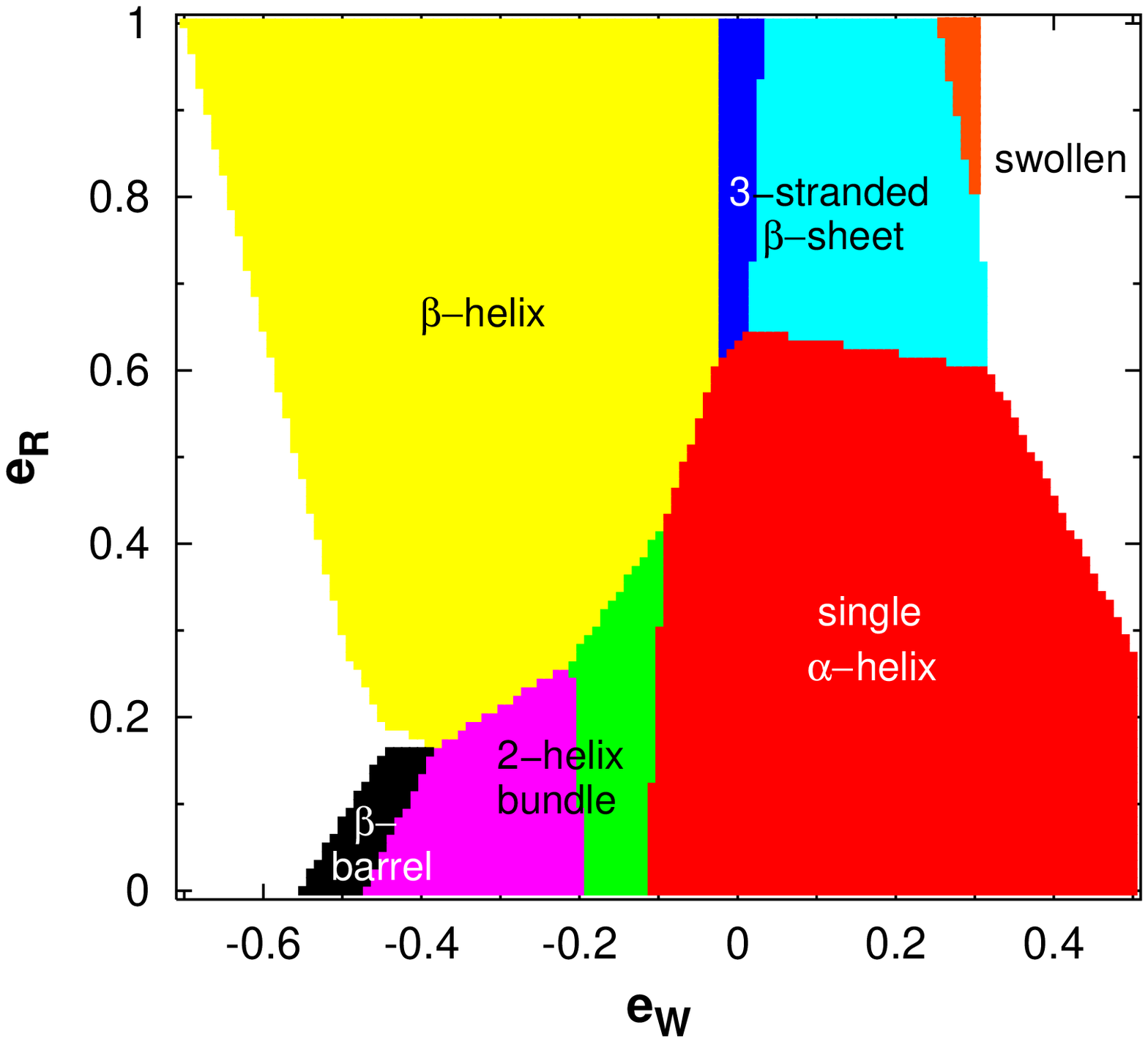}
%\caption{}
\end{figure}
\centerline{Figure-3 (Banavar)}

\newpage
\pagestyle{empty}

\begin{figure}
\centering
\includegraphics[width=10.4cm]{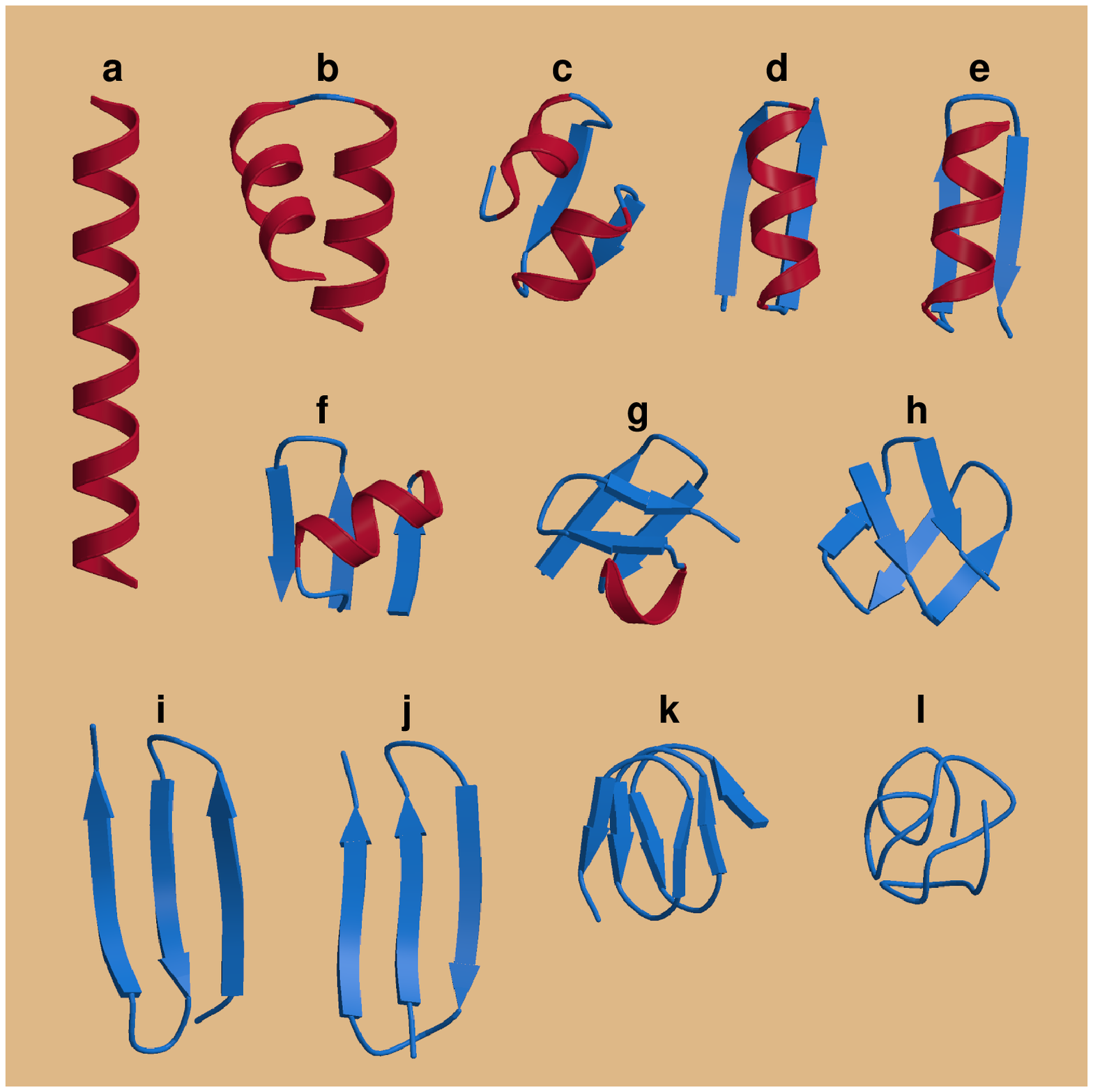}
%\caption{}
\end{figure}
\centerline{Figure-4 (Banavar)}

\newpage
\pagestyle{empty}

\begin{figure}
\centering
\includegraphics[width=10.4cm]{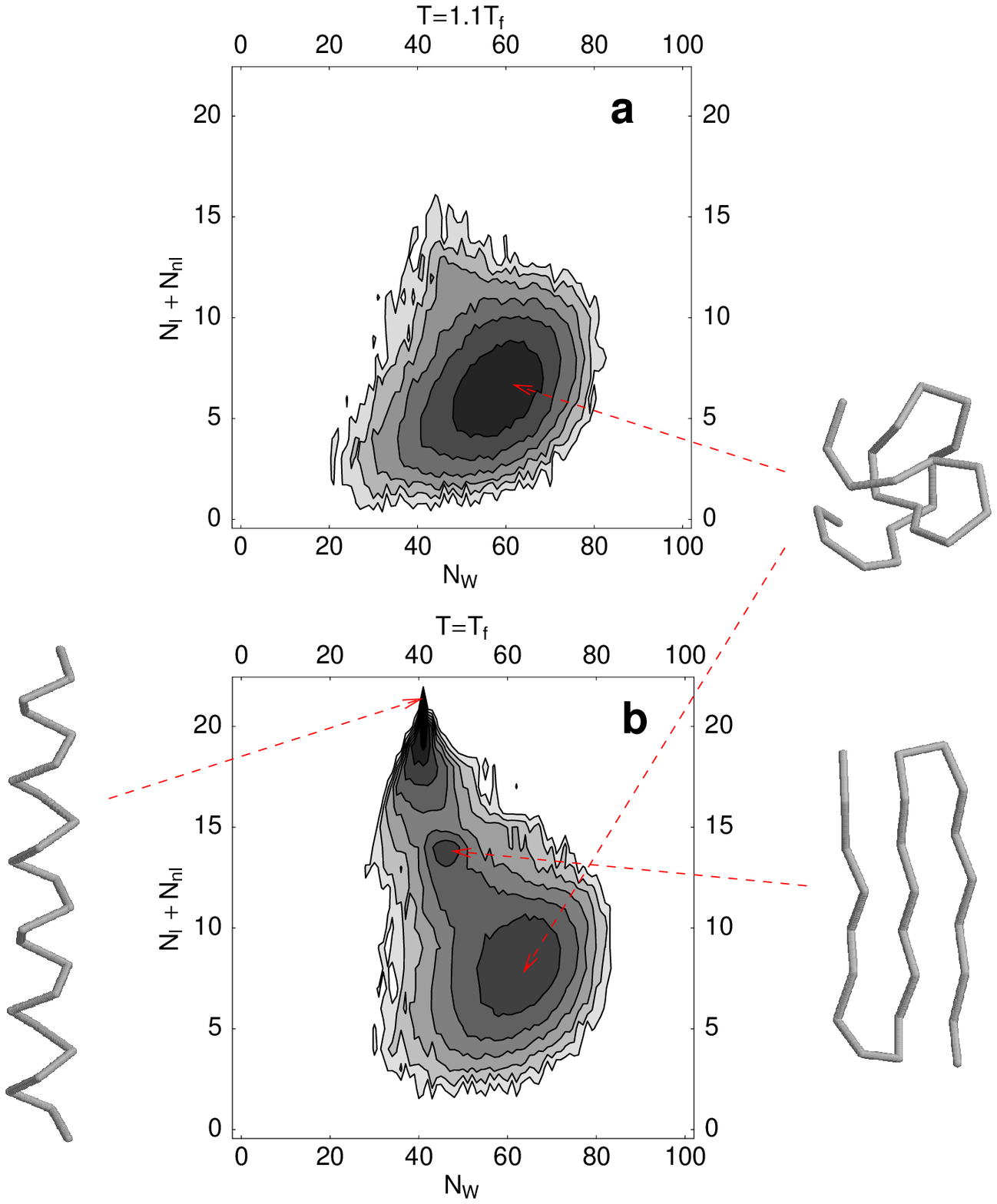}
%\caption{}
\end{figure}
\centerline{Figure-5 (Banavar)}

\end{document}